\documentclass[letterpaper, 10 pt, conference]{ieeeconf}  

\IEEEoverridecommandlockouts                              

\usepackage{cite}
\usepackage{amsmath,amssymb,amsfonts,bm}
\usepackage{algorithmic}
\usepackage{graphicx}
\usepackage{textcomp}
\usepackage{epsfig}
\usepackage{times}
\usepackage{longtable}
\usepackage{supertabular,booktabs}
\usepackage{ltablex}
\usepackage{array}
\usepackage{setspace}
\usepackage{xcolor,soul}
\usepackage{balance}
\usepackage{etoolbox}
\usepackage[labelsep=period,font={small}]{caption}

\newcolumntype{C}[1]{>{\centering\arraybackslash} m{#1}}

\newcommand{\qed}{\hfill $\square$}

\renewcommand{\thetable}{\arabic{table}}

\title{\LARGE \bf
MCE-based Direct FTC Method for Dynamic Positioning of Underwater Vehicles with Thruster Redundancy*
}

\author{Ji-Hong Li$^{1}$
\thanks{*This work was partially supported by the Korea Institute of Marine Science \& Technology Promotion (KIMST) funded by the Ministry of Ocean and Fisheries (RS-2024-00432366), also in part by the same organization with the grant No. RS-2023-00256122, all in the Republic of Korea.}
\thanks{$^{1}$Ji-Hong Li is with the Autonomous Systems R\&D Division, Korea Institute of Robotics and Technology Convergence,
        Jigok-Ro 39, Nam-Gu, Pohang 37666, Republic of Korea
        {\tt\small jhli5@kiro.re.kr}}%
}

\begin{document}

\maketitle
\thispagestyle{empty}
\pagestyle{empty}

\begin{abstract}
This paper presents an active model-based FTC (fault-tolerant control) method for the dynamic positioning of a class of underwater vehicles with thruster redundancy. Compared to the widely used state and parameter estimation methods, this proposed scheme directly utilizes the vehicle's motion control error (MCE) to construct a residual for detecting thruster faults and failures in the steady state of the control system. In the case of thruster fault identification, the most difficult aspect is that the actual control input with thruster faults is unknown. However, through a detailed and precise analyses of MCE variation trends in the case of thruster faults, highly useful information about this unknown control input can be extracted. This characteristic also serves as the foundation for the novel scheme proposed in this paper. As for control reconfiguration, it is straightforward since the thrust losses can be directly estimated as a result of the identification process. Numerical studies with the real world vehicle model are also carried out to demonstrate the effectiveness of the proposed method.
\end{abstract}

\section{Introduction}
Over the past few decades, interest in deep sea exploration has steadily grown, driven by its wide range of applications, including scientific research, resource exploration, oil and gas drilling, rescue operation, and various military activities. This insistent interest has fueled rapid advancements in underwater vehicle technology, further intensifying human curiosity about the deep sea.

Traditional ocean exploration typically consists of two consecutive steps of operations. The first one is a large-area survey conducted using torpedo-type autonomous underwater vehicles (AUVs), which are equipped with various sonar devices to inspect the seafloor environment. Once the AUV is recovered, recorded sonar images are manually analyzed to identify specific points of interest. The process then moves to the second step, where remotely operated vehicles (ROVs) or manned submersibles are deployed to these identified locations for detailed exploration and manipulation tasks, ultimately fulfilling the mission.

To ensure the reliability and safety of these expensive underwater robotic systems, ROVs and manned submersibles are typically designed with redundant thruster configurations. From a system science perspective, one of the most critical failure modes for these vehicles is thruster malfunction, alongside communication loss. As a result, the design of fault-tolerant control (FTC) strategies to address thruster faults has emerged as a highly attractive research topic in robotics and control engineering.

This paper focuses on the model-based active FTC problem for a class of underwater vehicles with thruster redundancy. Compared to the related previous works, the main contributions of this work can be summarized as follows:
\begin{itemize}
\item The primary goal of thruster FTC is to ensure proper vehicle motion control. With this in mind, in this paper, motion control error (MCE) is directly applied to construct a residual for detecting fault and failure. This approach is significantly more compact compared to the previous works, which rely on estimation errors to generate residuals and therefore require an additional estimation scheme.
\item A major challenge in thruster fault identification is the lack of knowledge about the actual control input when a fault occurs. However, through a detailed analysis of MCE variations under thruster faults, we demonstrate that valuable information about this unknown control input can be inferred—or even estimated. This insight plays a crucial role in the proposed fault identification scheme.
\item Once thruster faults are identified through the proposed method, control reconfiguration or allocation becomes straightforward, as these corresponding thrust losses can be directly estimated as part of the identification process.
\end{itemize}

\section{Related Works}
Fault detection and identification (FDI) methods based on analytical redundancy have been widely studied in various complex dynamics systems \cite{b1,b2}. For some of large scale systems where dynamic modeling is prohibitively time-consuming, model-free method, especially using the expert system and artificial neural network, is suitable for their FTC \cite{b3,b2}. However, this method usually requires a suitable training set data, which is difficult to be acquired for many real-world dynamics systems. In addition, training neural networks, whether in off-line \cite{b4} or on-line \cite{b5} learning cases, can be computationally intensive and time-consuming \cite{b2}. In contrast, model-based method leverages the idea of generating residuals that reflect the inconsistency between actual and estimated behavior \cite{b2,b6}. State-estimation and parameter-estimation methods are two of most common schemes in this group \cite{b7}--\hspace{1sp}\cite{b9}. These methods apply various system identification techniques to estimate system states or model parameters and using corresponding estimation errors to construct proper residuals to detect and diagnosis the faults and failures.

This paper addresses the model-based active FTC problem for a class of underwater vehicles with thruster redundancy. Quite a number of related works have been done thus far \cite{b10}--\hspace{1sp}\cite{b15}. In \cite{b11}, state estimation error with all states measurable is used to detect the fault, and further an unknown input-observer is proposed to isolate this fault. In \cite{b12}--\hspace{1sp}\cite{b14}, a sliding-mode observer is used to estimate unmeasurable states and use the deviation of sliding surface of this observer in the steady state to detect and isolate the thruster fault and failure. For control allocation or reconfiguration, all the weight parameters for each thruster are adjusted iteratively until the residual minimization is ensured.

For most of practical underwater vehicles, especially expensive work-class ROVs and manned submersibles, a precise and reliable navigation system is the cornerstone of the entire vehicle system. These vehicles are usually equipped with various navigation systems (see \cite{b16} and references therein) combining with precious acoustic positioning system and navigation sensors, such as IMU (inertial measurement unit) and DVL (Doppler velocity log). Therefore, from a control engineering perspective, there is no need for additional observers to estimate the vehicle's states. On the other hand, in the case of redundant system, it's convenient to design a proper control method to guarantee the various performances in practice \cite{b17,b18}. Building on these two perspectives, in the authors' previous work \cite{b19}, a novel scheme of thruster FTC for a class of underwater vehicles with thruster redundancy is proposed. Instead of relying on estimation errors, the scheme directly utilizes the vehicle's trajectory tracking error to construct a residual for detecting thruster faults and failures. For further fault identification, a novel diagnosis scheme is presented based on a detailed analysis of tracking error changing trends in response to variations in each thrust force in the steady state of control system. A key advantage of this approach is its ability to quickly and effectively pinpoint the faulty thruster by analyzing the trends of specific error combinations, such as 2D position and azimuth errors in the horizontal plane. As for control reconfiguration, the process is straightforward: the thrust loss parameter associated with the identified faulty thruster is adjusted until the residual falls below a specified threshold again.

This paper aims to extend the results from \cite{b19} to the case of dynamic positioning, with the following newly added contributions:
\begin{itemize}
\item Unlike \cite{b19}, where no uncertainty term is considered in the vehicle's dynamics, this paper accounts for unknown bounded uncertainty terms, including significant sea currents, for its dynamic positioning. All these uncertainties are assumed to be unmatched.
\item In \cite{b19}, the residual after the thruster fault remains elevated until the thrust loss parameter is adjusted to the correct value. In contrast, after the initial spike at the moment of thruster fault occurrence, the residual in this paper automatically decreases below the specified fault detection criterion. For this reason, the method in \cite{b19} is not directly applicable to the case of dynamic positioning considered in this paper. However, this automatic decrease in residual also reveals an important property: the actual control input after thruster fault remains the same or very similar to the control input before the fault. Since this input matches the designed control input, it is known to us. Using this estimated actual control input after thruster fault, we can easily formulate the identification method and thrust loss estimation process. And this represents a key contribution of this paper.
\item Furthermore, with this formulated identification method, the restrictive condition of \emph{Assumption 3} in \cite{b19}, which states that ``at each time, only one thruster occurs fault or failure,'' is removed in this paper. The proposed method in this paper can identify up to two simultaneous double thruster faults and can accurately estimate the thrust losses of each of them.
\end{itemize}

\section{Methodology}
\subsection{Problem Statement}
\subsubsection{Vehicle Model}
\begin{figure}[!t]
\centerline{\includegraphics[width=6.0cm]{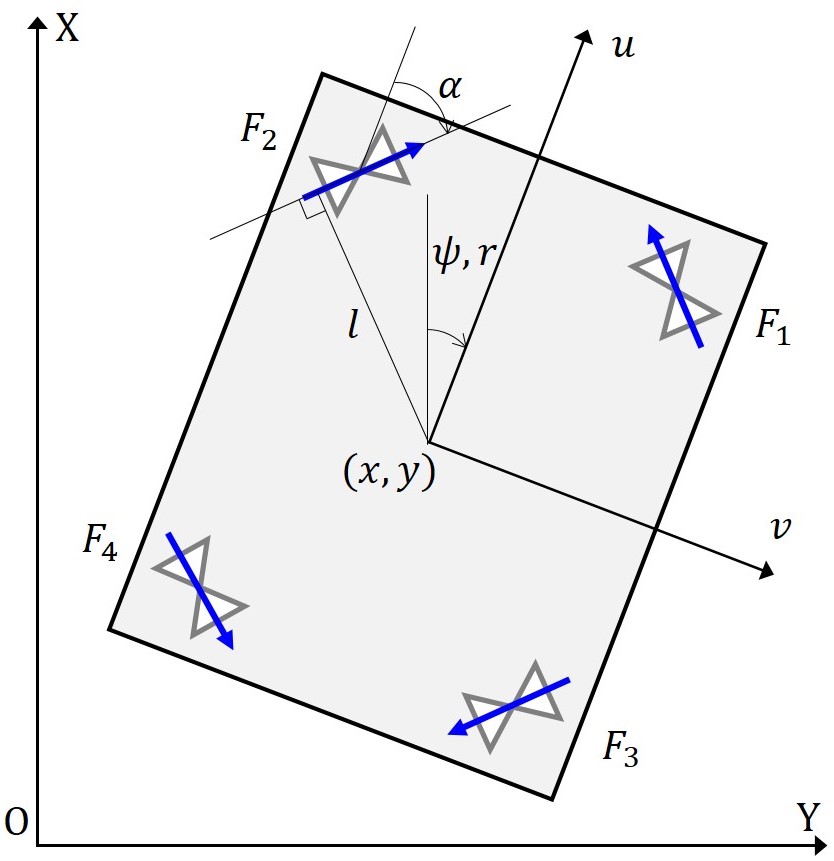}}
\caption{Vehicle configuration in the navigation frame and its thruster configuration in the body-fixed frame.}
\label{fig1}
\end{figure}

For most underwater vehicles, especially for work-class ROVs and manned submersibles, the center of gravity is often designed to be (much) lower than the center of buoyancy to stabilize roll motion. In addition, the thruster system is typically configured to allow easy decoupling of the vehicle's horizontal and vertical motions, although vertical thrusters are sometimes used for tilt control. Considering these aspects and for the sake of discussion, this paper focuses solely on the vehicle's 3DOF horizontal motion, with its kinematics and dynamics presented as follows \cite{b20}.
\begin{align}
\dot{\bm{\eta}}&=\bm{J\nu}, \label{eq1} \\
\bm{M}\dot{\bm{\nu}}+\bm{C\nu}+\bm{D\nu}+\bm{g}(\bm{\eta})&=\bm{\tau}+\bm{d}, \label{eq2}
\end{align}
where $\bm{\eta}=[x,y,\psi]^T$ denotes the vehicle configuration on the horizontal plane in the navigation frame, and $\bm{\nu}=[u,v,r]^T$ is the velocity vector in the vehicle's body-fixed frame, see Fig. \ref{fig1}; $\bm{M}$ indicates the mass term including added mass, $\bm{C}$ denotes the Coriolis and centripetal term, $\bm{D}$ is the damping components, and $\bm{g}$ presents gravitational term, all of which satisfy to be of class $C^2$; $\bm{\tau}=[\tau_u,\tau_v,\tau_r]^T$ is the thrust force and moment vector; $\bm{d}$ denotes the uncertainty term including modeling errors, measurement noises, exogenous disturbances such as sea currents, etc., all of which are unmatched; $\bm{J}$ denotes the coordinate transformation matrix from the vehicle's body-fixed frame to the navigation frame, and can be expresses as follows,
\begin{equation*}
\bm{J}=\begin{bmatrix}\cos\psi&-\sin\psi&0\\ \sin\psi&\cos\psi&0\\0&0&1\end{bmatrix}.
\end{equation*}

For uncertainty term $\bm{d}$, we have the following assumption.
\emph{Assumption 1}. $|d_i|\leq D_{mi},~i=1,2,3$ with $D_{mi}>0$ unknown constant.

Considering (\ref{eq2}), the vehicle's dynamics usually satisfies the following properties \cite{b20,b21}
\begin{verse}
$\textbf{P}1$. $\bm{M}=\bm{M}^T>0$, and $\dot{\bm{M}}(t)=0,~\forall t>0$. \\
$\textbf{P}2$. $\bm{s}^T\left[\dot{\bm{M}}-2\bm{C}\right]\bm{s}=0,~\forall \bm{s}\in\bm{\Re}^3$.\\
$\textbf{P}3$. $\bm{D}>0$.
\end{verse}

Unlike previous related works \cite{b12}--\hspace{1sp}\cite{b14}, where the control laws are derived in the navigation frame, this paper directly solves the control problem in the vehicle's body-fixed frame as in (\ref{eq2}).

For thruster configuration, we consider the case as shown in Fig. \ref{fig1}. This is also the most common configuration for underwater vehicles, where the arrangement of four horizontal thrusters has a symmetrical structure in the fore-aft and left-right directions. In this case, the control input $\bm{\tau}=[\tau_u,\tau_v,\tau_r]^T$ can be presented as follows:
\begin{align}
\bm{\tau}&=\bm{T}_{conf}\bm{F} \nonumber \\
&=\begin{bmatrix}\cos\alpha&\cos\alpha&-\cos\alpha &-\cos\alpha\\ -\sin\alpha&\sin\alpha&-\sin\alpha&\sin\alpha\\-l&l&l&-l\end{bmatrix}\begin{bmatrix}F_1\\F_2\\F_3\\F_4\end{bmatrix}, \label{eq3}
\end{align}
where $\alpha$ is the thruster orientation in the body-fixed frame and $l$ denotes the distance from the center point to the thruster's orientation line.

\subsubsection{Thruster Fault and Failure Model}
For each thruster, its thrust output depends on the specific thruster model,
\begin{equation}
F_i=F_T(u_i), \label{eq4}
\end{equation}
where $u_i$ is the thruster actual control input. Indeed, this is the input signal set into the thruster control board. In the case of BlueROV2 Heavy ROV \cite{b22}, whose model will be used in the numerical studies in this paper, the thruster model (\ref{eq4}) can be embodied as
\begin{equation}
F_i=K_iu_i, \label{eq5}
\end{equation}
with $K_i$ the thrust coefficient. Here the sign of $u_i$ indicates the thrust force direction: If $u_i>0$, the thruster outputs forward thrust; otherwise, reverse thrust.

For thruster fault and failure model, this paper directly implements the same model used in \cite{b10,b11} as follows:
\begin{equation}
F_i=K_iW_iu_i, \label{eq6}
\end{equation}
where $W_i$ is a numerical weight indicating the thrust loss and defined as follows,
\begin{equation}
W_i=\left\{\begin{matrix}0&\!~~~~~\text{if $i$-th thruster has a failure}\\0<W_i<1\!&\!\text{if $i$-th thruster is faulty}\\1&~\text{if $i$-th thruster is normal}\end{matrix}\right. \label{eq7}
\end{equation}

Consequently, the thruster force vector $\bm{F}$ can be expressed as follows:
\begin{equation}
\bm{F}=\bm{KWu}, \label{eq8}
\end{equation}
where $\bm{K}=diag(K_1,\cdots,K_4)$, $\bm{W}=diag(W_1,\cdots,W_4)$, and $\bm{u}=[u_1,\cdots,u_4]^T$.

Suppose $\bm{\tau}_c$ is the control input calculated by the controller, then control allocation \cite{b23} is to allocate the actual control input $\bm{u}$ as follows:
\begin{equation}
\bm{u}=\bm{W}^{-1}\bm{K}^{-1}\bm{T}_{conf}^{\dag}\bm{\tau}_c, \label{eq9}
\end{equation}
where superscript $``\dag"$ denotes the pseudo-inverse of a non-square matrix.

Here the problem is that, in practice, the exact value of $W$ is unknown, and therefore the control allocation has to take the following form
\begin{equation}
\hat{\bm{u}}=\hat{\bm{W}}^{-1}\bm{K}^{-1}\bm{T}_{conf}^{\dag}\bm{\tau}_c, \label{eq10}
\end{equation}
where $\hat{\bm{W}}$ is the estimation of $\bm{W}$ and $\hat{\bm{u}}$ is the corresponding control allocation.

Consequently, the embodied control input $\bm{\tau}$ becomes
\begin{equation}
\bm{\tau}=\bm{T}_{conf}\bm{KW}\hat{\bm{W}}^{-1}\bm{K}^{-1}\bm{T}_{conf}^{\dag}\bm{\tau}_c. \label{eq11}
\end{equation}

If $\hat{\bm{W}}=\bm{W}$, then we have $\bm{\tau}=\bm{\tau_c}$. Otherwise, the deviation of $\hat{\bm{W}}$ from $\bm{W}$ can cause nonzero $\Delta\bm{\tau}=\bm{\tau}_c-\bm{\tau}$, which further result in a similar deviation in the MCE during the steady state of the control system.

\subsubsection{Problem Formulation}
As in \cite{b19}, this paper aims to answer the question, ``\emph{Is it possible to detect the change in $\bm{W}$ and further identify which $W_i$ is changed, by solely observing the deviation of MCE during the steady state of the control system?}''

To address this problem, this paper supposes the following constraint conditions.

\emph{Assumption 2}. Thruster fault and failure occur only after the control system has been stabilized.

\emph{Assumption 3}. Thruster fault and failure is caused only by the reduction of corresponding numerical scale weight $\bm{W}$.

Compared to \cite{b19}, we can see that Assumption 3 in \cite{b19}, which might be the most restricting condition in practical applications, is eliminated in this paper.

\subsection{Motion Control Design}
Let $\bm{\eta}_d(t)$ denotes the vehicle's reference configuration and corresponding MCE is defined as $\bm{e}_{\eta}(t)=||\bm{\eta}_d(t)-\bm{\eta}(t)||_2$. Here, if $\dot{\bm{\eta}}_d=0$, then this is regulation control problem, otherwise, becomes trajectory tracking problem. The control objective in this paper is to design a control law for $\bm{\tau}$ so that $\bm{e}_{\eta}(t)$ to be uniformly ultimately bounded (UUB) with $t\rightarrow \infty$. For this purpose, the vehicle's kinematics and dynamics can be rewritten as follows:
\begin{align}
\dot{\bm{e}}_{\eta}&=\bm{\eta}_d-\bm{J\nu}, \label{eq12} \\
\bm{M}\dot{\bm{\nu}}&=-\bm{C\nu}-\bm{D\nu}-\bm{g}(\bm{\eta})+\bm{\tau}+\bm{d}. \label{eq13}
\end{align}

For this second-order strict-feedback form of motion control system, general backstepping method \cite{b24} is utilized to solve the control problem.

\subsubsection{Step 1}
For given MCE $\bm{e}_{\eta}(t)$, the following Lyapunov function candidate is considered in this step,
\begin{equation}
\bm{V}_1=\dfrac{1}{2}\bm{e}_{\eta}^T\bm{\Gamma}_1\bm{e}_{\eta}, \label{eq14}
\end{equation}
where $\bm{\Gamma}_1>0$ is a diagonal weighting matrix.

By differentiating (\ref{eq14}) and substituting (\ref{eq13}) into it, we get
\begin{equation}
\dot{\bm{V}}_1=\bm{e}_{\eta}^T\bm{\Gamma}_1\left(\dot{\bm{\eta}}_d-\bm{J\nu}\right), \label{eq15}
\end{equation}
according to which, the following control law is chosen for $\bm{\alpha}_{\nu}$, which is the stabilization function \cite{b24} for virtual control input $\bm{\nu}$ in (\ref{eq15}),
\begin{equation}
\bm{\alpha}_{\nu}=\bm{J}^{-1}\left(\dot{\bm{\eta}}_d+\bm{\Gamma}_1^{-1}\bm{A}_1\bm{e}_{\eta}\right), \label{eq16}
\end{equation}
where $\bm{A}_1>0$ is diagonal control gain matrix.

By substituting (\ref{eq16}) into (\ref{eq15}), we have
\begin{equation}
\dot{\bm{V}}_1=-\bm{e}_{\eta}^T\bm{A}_1\bm{e}_{\eta}+\bm{e}_{\eta}^T\bm{\Gamma}_1\bm{J}\bm{e}_{\nu}, \label{eq17}
\end{equation}
where $\bm{e}_{\nu}=\bm{\alpha}_{\nu}-\bm{\nu}=[e_u,e_v,e_r]^T$.

\subsubsection{Step 2}
Considering the vehicle's dynamics as (\ref{eq13}) with properties \textbf{P}1--\textbf{P}3, the following Lyapunov function candidate is taken in this step.
\begin{equation}
\bm{V}_2=\bm{V}_1+\dfrac{1}{2}\left[\bm{e}_{\nu}^T\bm{M}\bm{e}_{\nu}+\tilde{\bm{D}}_m^T\bm{\Gamma}_2\tilde{\bm{D}}_m\right], \label{eq18}
\end{equation}
where $\tilde{\bm{D}}_m=\hat{\bm{D}}_m-\bm{D}_m$ with $\bm{D}_m=[D_{m1},D_{m2},D_{m3}]^T$ and $\hat{\bm{D}}_m$ the estimation of $\bm{D}_m$, and $\bm{\Gamma}_2>0$ is a design diagonal weighting matrix.

By differentiating (\ref{eq18}) and substituting (\ref{eq17}) and (\ref{eq13}) into it, we have
\begin{align}
\dot{\bm{V}}_2=&-\bm{e}_{\eta}^T\bm{A}_1\bm{e}_{\eta}+\bm{e}_{\eta}^T\bm{\Gamma}_1\bm{J}\bm{e}_{\nu}+\dfrac{1}{2}\bm{e}_{\nu}^T\dot{\bm{M}}\bm{e}_{\nu} \nonumber \\
&+\bm{e}_{\nu}^T\left[\bm{M}\dot{\bm{\alpha}}_{\nu}+(\bm{C}+\bm{D})\bm{\nu}+\bm{g}(\bm{\nu})-\bm\tau-\bm{d}\right] \nonumber \\
&+\tilde{\bm{D}}_m^T\bm{\Gamma}_2\dot{\hat{\bm{D}}}_m. \label{eq19}
\end{align}

According to (\ref{eq19}), we choose the control law for $\bm{\tau}$ and corresponding adaptation law for $\hat{\bm{D}}_m$ as follows:
\begin{align}
\bm{\tau}&\!=\!\bm{A}_2\bm{e}_{\nu}\!\!+\!\!\bm{\Gamma}_1\bm{J}\bm{e}_{\eta}\!\!+\!\!\bm{M}\dot{\bm{\alpha}}_{\nu}\!\!+\!\!(\bm{C}\!\!+\!\!\bm{D})\bm{\alpha}_{\nu}\!\! +\!\!\bm{g}(\bm{\eta})\!\!+\!\!\hat{\bm{d}}, \label{eq20} \\
\dot{\hat{\bm{D}}}_m&\!=\!\bm{\Gamma}_2^{-1}\bm{e}_{\nu}, \label{eq21}
\end{align}
where $\bm{A}_2>0$ is diagonal control gain matrix, and $\hat{\bm{d}}\in\Re^3$ with $\hat{d_i}=\hat{D}_{mi}\varphi\left(\hat{D}_{mi}e_{\nu(i)}\right)$. Here, the smooth function $\varphi(\cdot)$ satisfies the following lemma \cite{b25}.

\emph{Lemma 1}. $\forall \epsilon>0$, there exist a smooth function $\varphi(\cdot)$ such that $\varphi(0)=0$ and
\begin{equation}
|\xi|\leq\xi\varphi(\xi)+\epsilon,~~\forall \xi\in\Re. \label{22}
\end{equation}

\emph{Remark 1}. One example of smooth function satisfying Lemma 1 is $\varphi(\xi)=tanh(\kappa\xi/\epsilon)$ with $\kappa$ the solution of the equation $\kappa=e^{-(\kappa+1)}$ \cite{b25}.

\emph{Theorem 1}. Consider the trajectory tracking problem for (\ref{eq1}) and (\ref{eq2}). If we choose the control law as (\ref{eq20}) and corresponding adaptation law as (\ref{eq21}), then we can guarantee the closed-loop control system to be UUB.

\emph{Proof}. By substituting (\ref{eq20}) and (\ref{eq21}) into (\ref{eq19}), we can get
\begin{align}
\dot{\bm{V}}_2=&-\bm{e}_{\eta}^T\bm{A}_1\bm{e}_{\eta}-\bm{e}_{\nu}^T\bm{A}_2\bm{e}_{\nu}+\dfrac{1}{2}\left(\dot{\bm{M}}-2\bm{C}\right)\bm{e}_{\nu} \nonumber \\
&-\bm{e}_{\nu}^T\bm{D}\bm{e}_{\nu}-\bm{e}_{\nu}^T\bm{d}-\bm{e}_{\nu}^T\hat{\bm{d}}+\tilde{\bm{D}}_m^T\bm{e}_{\nu} \nonumber \\
\leq & -\bm{e}_{\eta}^T\bm{A}_1\bm{e}_{\eta}-\bm{e}_{\nu}^T\bm{A}_2\bm{e}_{\nu}+\epsilon_1+\epsilon_2+\epsilon_3. \label{eq23}
\end{align}
In the final expansion, Lemma 1 is applied with $\epsilon_1,\epsilon_2,\epsilon_3>0$ the design parameters defined as in Lemma 1. (\ref{eq23}) indicates that both of $\bm{e}_{\eta}$ and $\bm{e}_{\nu}$ are all bounded by the constants $\epsilon_1,\epsilon_2,\epsilon_3>0$, and this concludes the proof. \qed

\emph{Remark 2}. Since $\epsilon_1,\epsilon_2,\epsilon_3>0$ can be chosen arbitrarily small, it's possible for us to get arbitrarily small value of MCE $\bm{e}_{\eta}$ and $\bm{e}_{\nu}$. However, too small value of $\epsilon_1$, $\epsilon_2$, and $\epsilon_3$ may cause certain high-gain control problem. Therefore, these parameters should be carefully chosen in practice.

\subsection{Fault Detection and Identification}
In practical applications, the thrusters are initially presumed to be functioning normally. In this case, we have $\hat{\bm{W}}=\bm{W}=diag(1,\cdots,1)$. Suppose there is a thrust loss $\Delta \bm{W}$ at $t=t_c>0$. In this case, $\bm{W}=\hat{\bm{W}}-\Delta \bm{W}$, combining with which, (\ref{eq11}) can be expanded as follows:
\begin{align}
\bm{\tau}&=\bm{T}_{conf}\bm{KW}\hat{\bm{W}}^{-1}\bm{K}^{-1}\bm{T}_{conf}^{\dag}\bm{\tau}_c \nonumber \\
&=\overbrace{\begin{bmatrix}\tau_1&\tau_2&\tau_3&\tau_4\\-\tau_1&\tau_2&\tau_3&-\tau_4\\-a\tau_1&a\tau_2&-a\tau_3&a\tau_4\end{bmatrix}}^{\textbf{\normalsize{$\bm{M}_{\tau}$}}}\begin{bmatrix}1-{\Delta W_1}/{\hat{W}_1}\\ \vdots\\1-{\Delta W_4}/{\hat{W}_4}\end{bmatrix} \nonumber \\
&=\bm{\tau}_c-\bm{M}_{\tau}\delta \bm{W}, \label{eq24}
\end{align}
where $\delta \bm{W}\in\Re^4$ with $\delta \bm{W}(i)=\Delta W_i/\hat{W}_i$, $i=1,\cdots,4$, and
\begin{equation}
\begin{bmatrix}\tau_1\\ \tau_2\\ \tau_3\\ \tau_4\end{bmatrix}=\begin{bmatrix}1&-1&-1/a\\1&1&1/a\\1&1&-1/a\\1&-1&1/a\end{bmatrix}\bm{\tau}_c=\bm{M}_c^T\bm{\tau}_c, \label{eq25}
\end{equation}
with $a=l\sec\alpha=0.267$. In the above expansion, for the sake of simplicity, we directly apply the parameters $\alpha=\pi/4$ and $l=0.1888m$ of BlueROV2 Heavy ROV model \cite{b22}.

\emph{Proposition 1}. If the designed controller $\bm{\tau}_c$ calculated as (\ref{eq20}) has no component equal to zero, then $\bm{M}_{\tau}$ has rank 3.

\emph{Proof}. Since $\bm{M}_c$ has rank 3, $\bm{\tau}_c$ with non component equal to zero can guarantee at least three of $\tau_i$, $i=1,\cdots,4$ are nonzero, which further guarantees $\bm{M}_{\tau}$ to be rank 3. \qed

\emph{Remark 3}. Here, it's worth to mention that all components of $\bm{\tau}_c$ are nonzero is a sort of sufficient condition, not necessary one. It's easy to see that any of $\bm{\tau}_c\in\{[\zeta,0,0]^T,[0,\zeta,0]^T,[0,0,\zeta]^T\}$ with $\zeta\neq 0$ can guarantee all $\tau_i$ to be nonzero, and therefore easy to guarantee the same result of $\bm{M}_{\tau}$ having rank 3.

\emph{Remark 4}. If $\bm{M}_{\tau}$ has rank 3, it is straightforward to verify that up to two simultaneous thruster faults can be identified. However, in the case of three simultaneous thruster faults, it is, in principle, impossible to distinguish this fault from the one where the remaining single thruster occurs fault.

\emph{Remark 5}. It is also worth noting that although no component of $\bm{\tau}_c$ is exactly zero, it still possible for one of $\tau_i$, $i=1,\cdots,4$, to be equal or near zero. In this case, from (\ref{eq24}), the corresponding column vector in $\bm{M}_{\tau}$ becomes one with components of zero or very small magnitude. Consequently, the thrust loss associated with this column vector may not be sufficiently reflected in the residual, making it undetectable. To the best of the authors' knowledge, such detection failure is expected to be challenging regardless of any additional estimation methods applied.

If $\bm{\tau}$ is known after $t>t_c$, then through (\ref{eq24}), it's possible for us to identify $\delta\bm{W}$. The problem is that, in practice, it's impossible to know $\bm{\tau}$ since $\bm{W}$ is unknown after $t>t_c$. To solve this problem, we carry out a simulation study using BlueROV2 model \cite{b22} and investigate the similar case. In the simulation, the control law (\ref{eq20}) is applied with the adaptation law taken as (\ref{eq21}). Other detailed simulation conditions will be illustrated in Section IV.

\begin{figure}[!t]
\centerline{\includegraphics[width=\columnwidth]{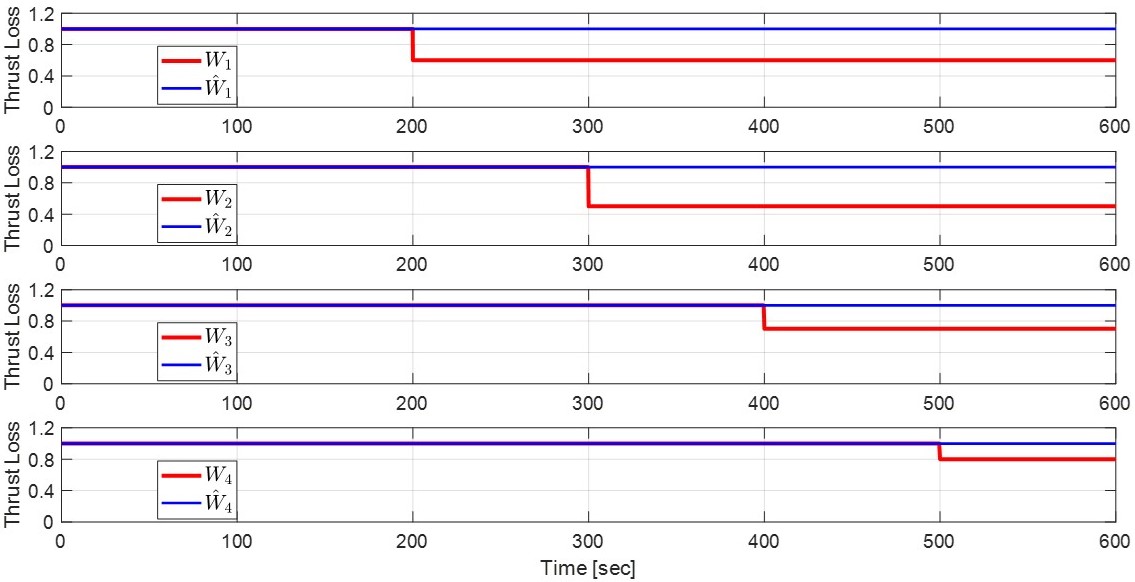}}
\caption{Thruster fault occurrence scenario.}
\label{fig2}
\end{figure}

\begin{figure}[!t]
\centerline{\includegraphics[width=\columnwidth]{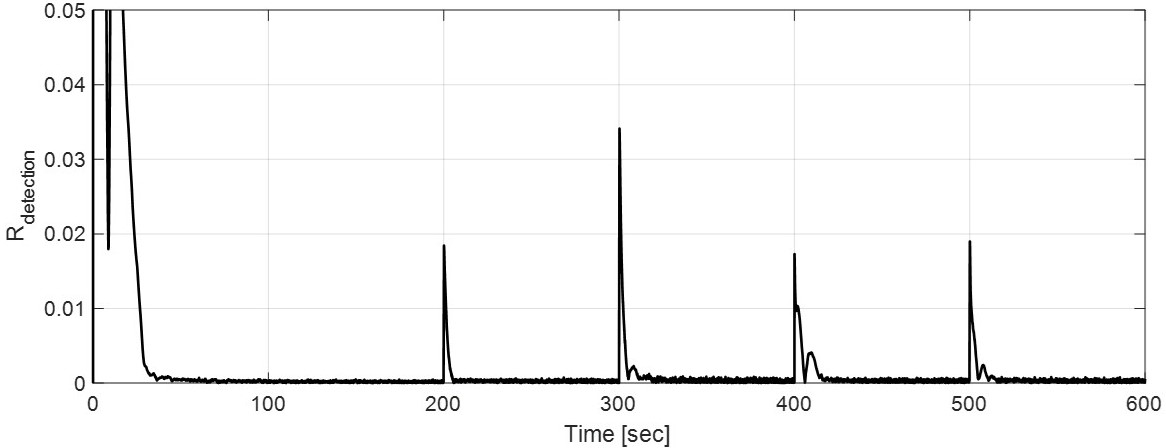}}
\caption{Corresponding residual measurement.}
\label{fig3}
\end{figure}

\begin{figure}[!t]
\centerline{\includegraphics[width=\columnwidth]{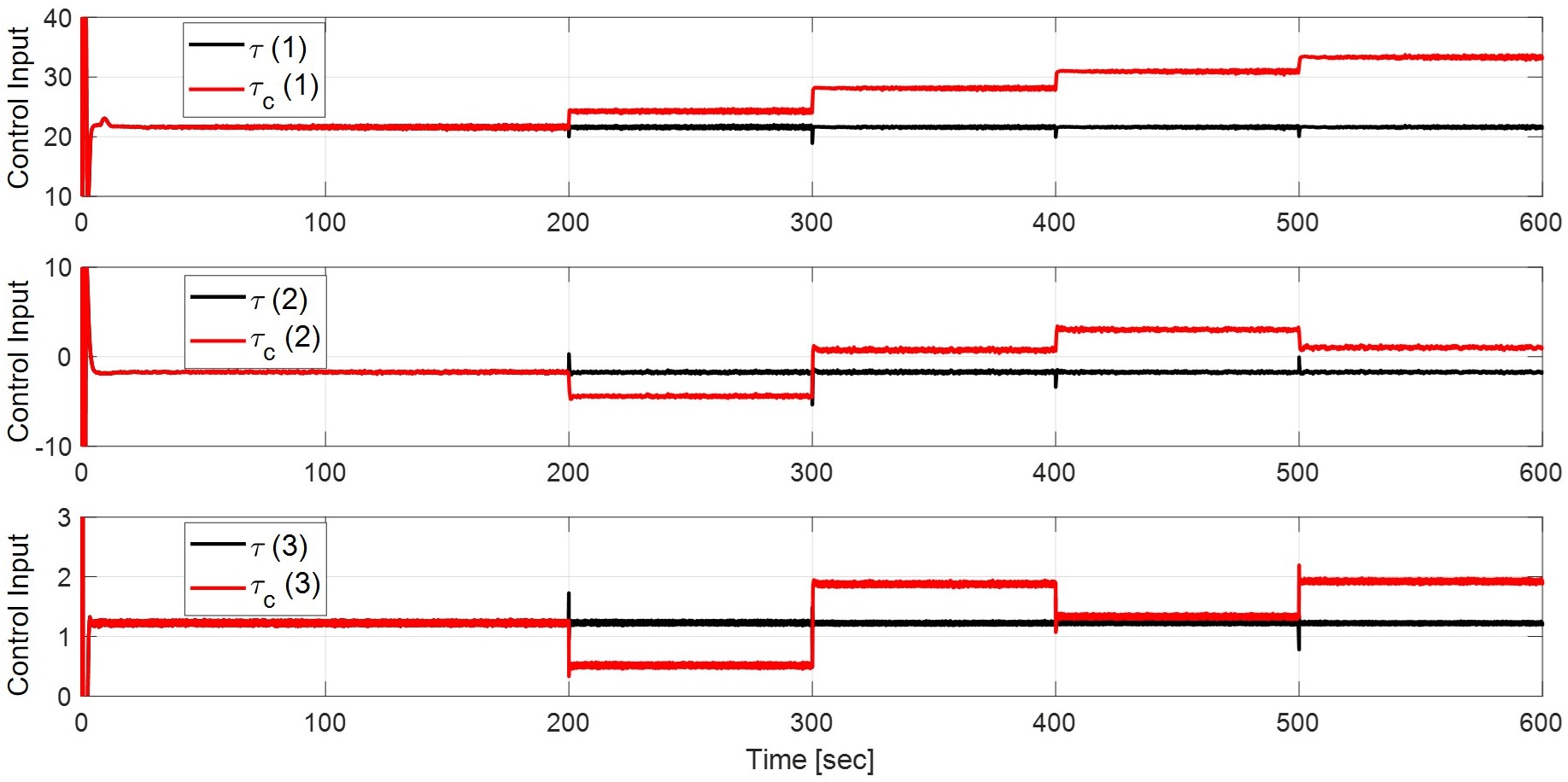}}
\caption{Corresponding time variations of designed controller $\bm{\tau}_c$ and actual control input $\bm{\tau}$.}
\label{fig4}
\end{figure}

Simulation results are shown in Fig. \ref{fig2}--\ref{fig4}. Fig. \ref{fig2} shows the sequential occurring of thruster faults, and corresponding residual changing trend is presented in Fig. \ref{fig3}, from which we can see that the constructed residual is suitable to detect the thruster faults. An additional point to note in Fig. \ref{fig3} is that, after sparking at the time of $\Delta \bm{W}$ occurrence, $R_{detection}$ automatically decreases again. This is a significantly different trend from the case in the authors' previous work \cite{b19}, where the residual stays elevated if $\hat{\bm{W}}$ remains unchanged. To the authors' knowledge, this kind of auto-decreasing of residual is due to the adaptation component $\hat{\bm{d}}$ in the control law (\ref{eq20}). Compared to the previous work \cite{b19}, this kind of auto-decreasing of residual imposes two challenges; one is making it difficult to evaluate the magnitude of $\Delta \bm{W}$, and the other is that the criterion for adjusting $\hat{\bm{W}}$ disappears. So we need to explore a new criterion for fault identification that can replace the residual in \cite{b19}. Fortunately, from Fig. \ref{fig4}, we can see that: before thruster faults occurrence ($t<200s$), $\bm{\tau}_c$ is exactly the same as $\bm{\tau}$, and along with the sequential occurrences of thruster faults, compared to $\bm{\tau}_c$, which correspondingly changes over the faults, $\bm{\tau}$ always maintains its original value as in $t<200s$ (except for the sparking at the very moment of fault occurrences). This also coincides with the result in Fig. \ref{fig3}. To conclude, the actual control input $\bm{\tau}$ after thruster faults occurrence can be approximated as $\bm{\tau}_c$ in the steady state case before thruster fault. Therefore, (\ref{eq24}) can be rewritten as
\begin{equation}
\bm{M}_{\tau}\delta \bm{W}=\bm{\tau}_c\mid_{t>t_c}-\bm{\tau}_c\mid_{t<t_c}+\bm{w}, \label{eq26}
\end{equation}
where $\bm{w}=\bm{\tau}-\bm{\tau}_c\mid_{t>t_c}$.

In this paper, $\bm{w}$ is assumed to be a zero mean random noise. Since $\bm{M}_{\tau}$ and $\bm{\tau}_c$ are always known, now we can identify the thruster faults $\delta \bm{W}$.

\subsubsection{Fault Detection}
Residual $R_{detection}$ is defined as follows \cite{b19}:
\begin{equation}
R_{detection}=\sqrt{x_e^2+y_e^2+c_1\psi_e^2}, \label{eq27}
\end{equation}
where $c_1>0$ is a normalization constant parameter. At each time $t>0$, fault detection for the calculated residual $R_{detection}$ is simple and can be described as follows:
\begin{align*}
&\textbf{if}~~ (R_{detection}> c_2){\scriptstyle\&\&}(\text{bFirst}==1) \\
&~~~\text{faultTrig}=true; \\
&~~~\text{bFirst}=0; \\
&\textbf{end if}
\end{align*}
where $c_2>0$ is a constant criterion.

\subsubsection{Fault Identification}
Fault identification follows the detection phase with $faultTrig=true$, see Algorithm 1.
\begin{table}[ht]
\small
\centering
\tabcolsep=-0.0cm
\setlength{\arrayrulewidth}{1.4pt}
\begin{spacing}{1.1}
\begin{tabular}{@{}cl@{}} \hline
&\textbf{Algorithm 1:} Fault Identification ( ) \\ \hline
&~\textbf{Input:} $\bm{M}_{\tau},\bm{\tau}_c,\hat{\bm{W}}$ \\
1&~\textbf{waiting for} faultTrig=true \\
2&~~~$dW_1=0$; $dW_2=[0;0]$; $nC=0$; \\
3&~~~\textbf{while} (faultTrig) \\
4&~~~~~\textbf{if} ($t-t_c>T_1$) \\
5&~~~~~~~[$\delta W_n,rmse,n]=tFaultID(\bm{M}_{\tau},\bm{\tau}_c,\hat{\bm{W}},single)$ \\
6&~~~~~~~\textbf{if} $rmse\leq c_3$ \\
7&~~~~~~~~~$dW_1{\scriptstyle+}=\delta W_n$; \\
8&~~~~~~~~~$nC{\scriptstyle++}$; \\
9&~~~~~~~~~\textbf{if} $t-t_c==T_2$ \\
10&~~~~~~~~~~~$\hat{W}(n,n)=\hat{W}(n,n)(1-dW_1/nC)$; \\
11&~~~~~~~~~\textbf{end if} \\
12&~~~~~~~\textbf{else} \\
13&~~~~~~~~~[$\delta W_{n_1},\delta W_{n_2},rmse,n_1,n_2]$= \\
&~~~~~~~~~~~~~~~~~~~~~~~~~~~~~~$tFaultID(\bm{M}_{\tau},\bm{\tau}_c,\hat{\bm{W}},double)$ \\
14&~~~~~~~~~\textbf{if} $rmse\leq c_4$ \\
15&~~~~~~~~~~~$dW_2{\scriptstyle+}=[\delta W_{n_1};\delta W_{n_2}]$; \\
16&~~~~~~~~~~~$nC{\scriptstyle++}$; \\
17&~~~~~~~~~~~\textbf{if} $t-t_c==T_2$ \\
18&~~~~~~~~~~~~~$\hat{W}(n_1,n_1)=\hat{W}(n_1,n_1)(1-dW_2[1]/nC)$; \\
19&~~~~~~~~~~~~~$\hat{W}(n_2,n_2)=\hat{W}(n_2,n_2)(1-dW_2[2]/nC)$;~~~ \\
20&~~~~~~~~~~~\textbf{end if} \\
21&~~~~~~~~~\textbf{end if} \\
22&~~~~~~~\textbf{end if} \\
23&~~~~~\textbf{end if} \\
24&~~~~~\textbf{if} $t-t_c==T_3$ \\
25&~~~~~~~\text{faultTrig}=false; \\
26&~~~~~~~\text{bFirst}=1; \\
27&~~~~~\textbf{end if} \\
28&~~~\textbf{end while} \\
29&~\textbf{end waiting for} \\
30&~\textbf{return} \\ \hline
\end{tabular}
\end{spacing}
\end{table}

In Algorithm 1, $\bm{\tau}_c$ represents both of $\bm{\tau}_c\mid_{t<t_c}$ and $\bm{\tau}_c\mid_{t>t_c}$; $c_3,c_4>0$ are criterion parameters; $T_3>T_2>T_1>t_c$ are time-related parameters, where $t_c$ represents the fault occurrence time. At each $t=t_c$, the designed control input $\bm{\tau}_c$ experiences magnitude jumps, and $T_1$ is used to allow for the convergence of this jumped $\bm{\tau}_c(t)$. Due to the presence of random noise $\bm{w}$ as in (\ref{eq26}), the estimated thrust loss $\delta W_i$ also contains random noise. To mitigate this effect, $T_2$ is used to attenuate the impact of this noise by simply averaging the estimation $\delta W_i$ over the period $T_2-T_1$. Each occurrence of control allocation at $t=t_c+T_2$ using updated thrust loss parameter $\hat{\bm{W}}$ is typically followed by a spike in $R_{detection}$. To account for this, $T_3$ is applied to allow for the re-convergence of $R_{detection}$ after $t=t_c+T_2$.

To illustrate the core function $tFaultID( )$ in Algorithm 1, we rewrite (\ref{eq26}) as follows:
\begin{equation}
\begin{bmatrix}\tau_1&\tau_2&\tau_3&\tau_4\\-\tau_1&\tau_2&\tau_3&-\tau_4\\-a\tau_1&a\tau_2&-a\tau_3&a\tau_4\end{bmatrix}\begin{bmatrix}\delta W_1\\ \vdots \\ \delta W_4\end{bmatrix}=\bm{b}_{\tau}, \label{eq28}
\end{equation}
where $\bm{b}_{\tau}=\bm{\tau}_c\mid_{t>t_c}-\bm{\tau}_c\mid_{t<t_c}$.

From (\ref{eq28}), we can see that in the case of single thruster fault, identification problem becomes to find out a vector among four column vectors of $\bm{M}_{\tau}$, which is parallel to $\bm{b}_{\tau}$ and points in the same direction as $\bm{b}_{\tau}$. Corresponding pseudo code for this process is shown in Algorithm 2, where $rmse$ means root mean square error.
\begin{table}[ht]
\small
\centering
\tabcolsep=-0.0cm
\setlength{\arrayrulewidth}{1.4pt}
\begin{spacing}{1.1}
\begin{tabular}{@{}cl@{}} \hline
&\textbf{Algorithm 2:} $tFaultID(\bm{M}_{\tau},\bm{\tau}_c,\hat{\bm{W}},single)$ \\ \hline
1&~\textbf{for} i=1:4 \\
2&~~~$\delta {W}[i]=\bm{M}_{\tau}^{\dag}(:,i)\bm{b}_{\tau}$; \\
3&~~~$rmse[i]=\sqrt{\dfrac{1}{3}\sum_{j=1}^{3}\left[b_{\tau}(j)-\delta W[i]M_{\tau}(j,i)\right]^2}$; \\
4&~\textbf{end for} \\
5&~\textbf{return} minimum $rmse[i]$ and corresponding $\delta W[i]$ and $i$ \\ \hline
\end{tabular}
\end{spacing}
\end{table}

In the case of double thruster faults, the identification process is similar to that in Algorithm 2, except that the column vector in Algorithm 2 is replaced by a combination of two column vectors from $\bm{M}_{\tau}$. Given four column vectors $\bm{M}_{\tau}(:,i),~i=1,\cdots,4$, there are a total of six different combinations of two columns. Let $\bm{\Omega}$ denotes a $3\times2$ matrix set consisting of these six different combinations. The identification process for double thruster faults can then be described as in Algorithm 3.
\begin{table}[ht]
\small
\centering
\tabcolsep=-0.0cm
\setlength{\arrayrulewidth}{1.4pt}
\begin{spacing}{1.1}
\begin{tabular}{@{}cl@{}} \hline
&\textbf{Algorithm 3:} $tFaultID(\bm{M}_{\tau},\bm{\tau}_c,\hat{\bm{W}},double)$ \\ \hline
1&~Construct $\bm{\Omega}$ from $\bm{M}_{\tau}$; \\
2&~\textbf{for} i=1:6 \\
3&~~~$\bm{a}_i\in\bm{\Omega}$; \\
4&~~~$\delta \bm{W}(:,i)=\bm{a}_i^{\dag}(:,i)\bm{b}_{\tau}$; \\
5&~~~$rmse[i]=\sqrt{\dfrac{1}{3}\sum_{j=1}^{3}\left[{b}_{\tau}(j)-\delta W(1,i)a_i(j,1)\right.}$ \\
&~~~~~~~~~~~~~~~~~~~$\overline{-\left.\delta W(2,i)a_i(j,2)\right]^2}$; \\
6&~\textbf{end for} \\
7&~\textbf{return} minimum $rmse[i]$ and corresponding $\delta \bm{W}(:,i)$ and $i$ \\ \hline
\end{tabular}
\end{spacing}
\end{table}

In Algorithm 3, from returned case number $i$, we can further decode the corresponding column numbers $j,k\in\{1,\cdots,4\}$ with $\bm{a}_i=[\bm{M}_{\tau}(:,j);\bm{M}_{\tau}(:,k)]$.

\subsubsection{Control Allocation}
Control allocation can be efficiently achieved using (\ref{eq10}), where $\hat{\bm{W}}$ is replaced with the estimated values as specified in Algorithm 1 (line 10 or line 18 and 19).

\section{Numerical Studies}
The 3DOF horizontal dynamic positioning of an underwater vehicle in a strong sea current environment is considered in this numerical studies. As mentioned before, the 6DOF nonlinear model of BlueROV2 Heavy ROV \cite{b22} is directly applied in the Matlab-based simulation. The sea current is simulated with $u_c=1.0m/s$ and $\psi_c=-120^o$, while the remaining uncertainty terms are set as $\bm{d}-\bm{d}_{sc}=[1-2rand;1-2rand;0.03(1-rand)]$ with $\bm{d}_{sc}$ indicating the aforementioned sea current. Here, $rand\in[0,1]$ denotes a random variable.

For the vehicle's thruster configuration, we have $\bm{K}=diag(40,40,40,40)$, $\alpha=45^o$ and $l=0.1888m$. The vehicle's dynamic positioning reference configuration is set as $\bm{\eta}_d=[10m,2m,70^o]^T$, and the vehicle's initial state is chosen as $\bm{X}(0)=\bm{0}^{12\times1}$ (in the simulation, 6DOF model is used instead of 3DOF). Controller design parameters are set as $\bm{\Gamma}_1=diag(1,1,1500)$, $\bm{\Gamma}_2=diag(1,1,0.6)$, $\bm{A}_1=diag(1.5,1.5,5)$, $\bm{A}_2=diag(1,1,12)$, $\hat{\bm{D}}_m=\bm{0}^{3\times1}$, and $\epsilon_1=\epsilon_2=\epsilon_3=0.1$. For the smooth function $\varphi(\cdot)$, it is chosen as $\varphi(\xi)=\tanh(\kappa \xi/\epsilon)$ in the simulation with $\kappa=0.2785$ \cite{b25}. FDI algorithm related parameters are chosen as: $c_1=(180/\pi)^2$, $c_2=0.005$, $c_3=c_4=0.1$, $T_1=20s$, $T_2=25s$, and $T_3=35s$. Finally, the algorithm sampling time is set as $0.01s$.

\begin{figure}[!t]
\centerline{\includegraphics[width=\columnwidth]{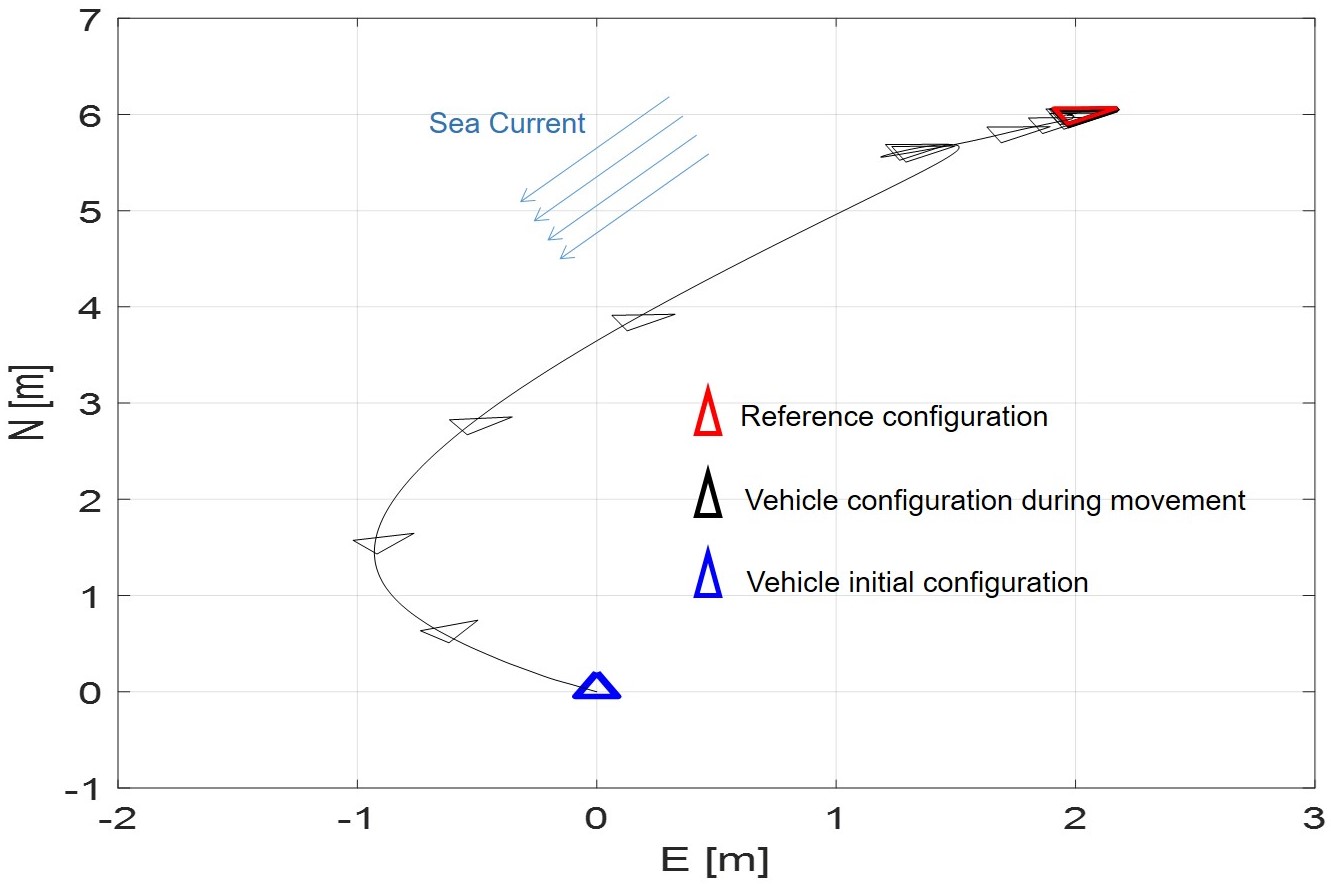}}
\caption{Vehicle trajectory during the dynamic positioning without thrust loss parameter update.}
\label{fig5}
\end{figure}

In the simulation, we carry out various case studies, among which Fig. \ref{fig2}--\ref{fig4} present a scenario where sequential thruster faults occur as $W_1\rightarrow W_4$, without control allocation based on thrust loss parameters estimation (line 10 in Algorithm 1 is not executed). Fig. \ref{fig5} shows the vehicle's trajectory for this case study, which closely resembles the trajectory observed when no thruster fault occurred. Some interesting results from other case studies are summarized below.

\subsubsection{A Single Thruster Fault with Sequential Thrust Loss Leading to Failure}
\begin{figure}[!t]
\centerline{\includegraphics[width=\columnwidth]{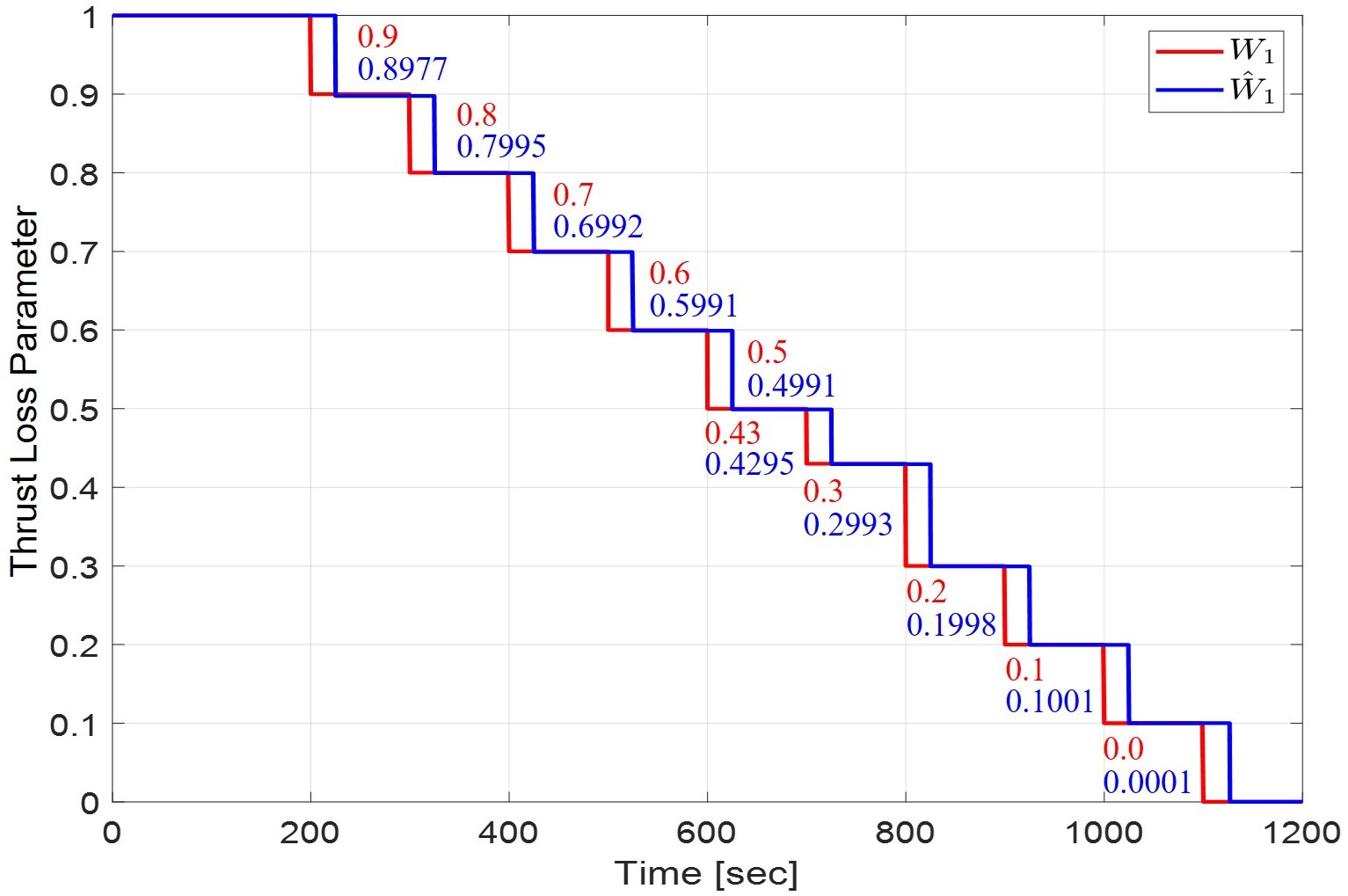}}
\caption{Single thruster sequential faults and their detection, identification, and corresponding thrust loss estimations.}
\label{fig6}
\end{figure}

\begin{figure}[!t]
\centerline{\includegraphics[width=\columnwidth]{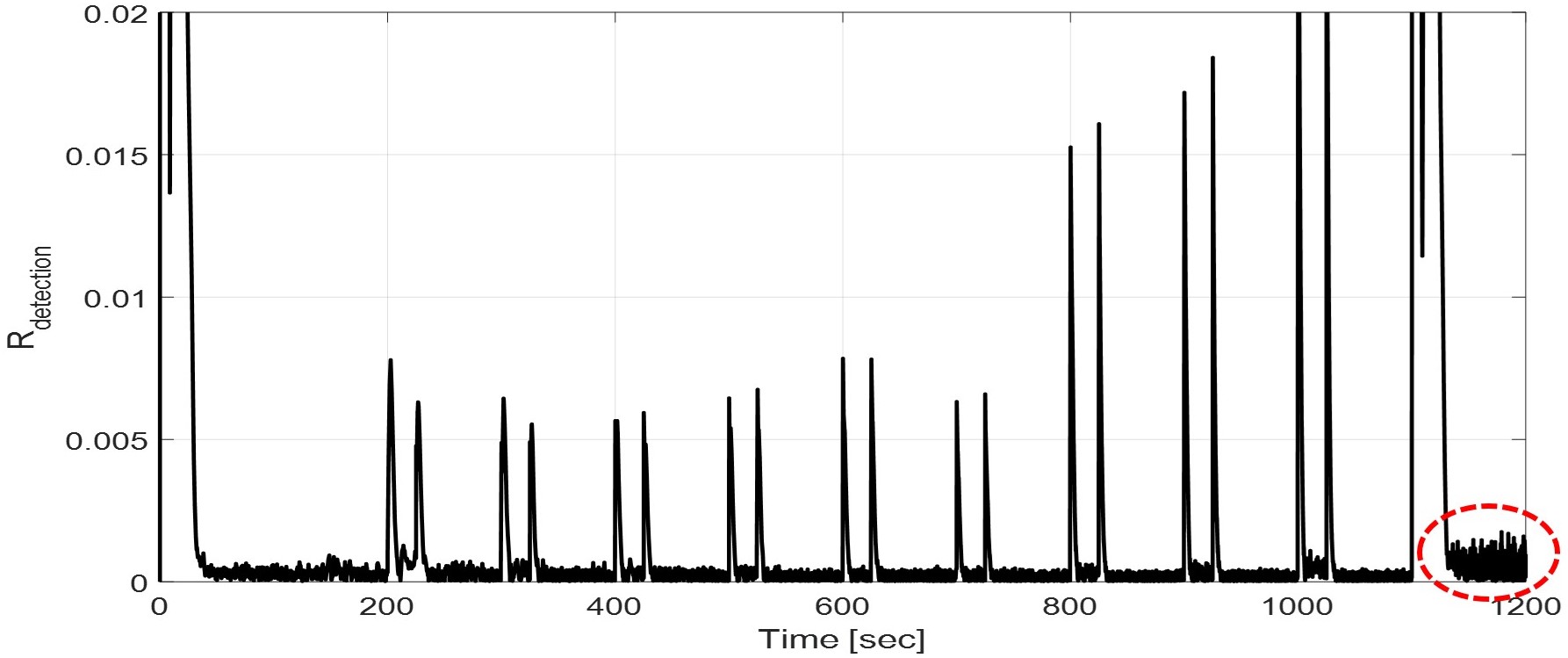}}
\caption{Residual variation trend during sequential single thruster faults leading to failure.}
\label{fig7}
\end{figure}

The simulation results for this case are presented in Fig. \ref{fig6} and \ref{fig7}. Fig. \ref{fig6} demonstrates that the proposed FDI algorithm effectively detects and identifies faults while maintaining accurate estimation capability. Notably, as shown in Fig. \ref{fig7}, following the single thruster failure at $t_c=1000s$, the residual exhibits a slight increase compared to the previous sequential stages, highlighted by the red-dotted circle. This suggests a discontinuity in control allocation at the moment when the thruster configuration transitions from four thrusters to three.

\subsubsection{Combination of Single and Double Thruster Faults}
\begin{figure}[!t]
\centerline{\includegraphics[width=\columnwidth]{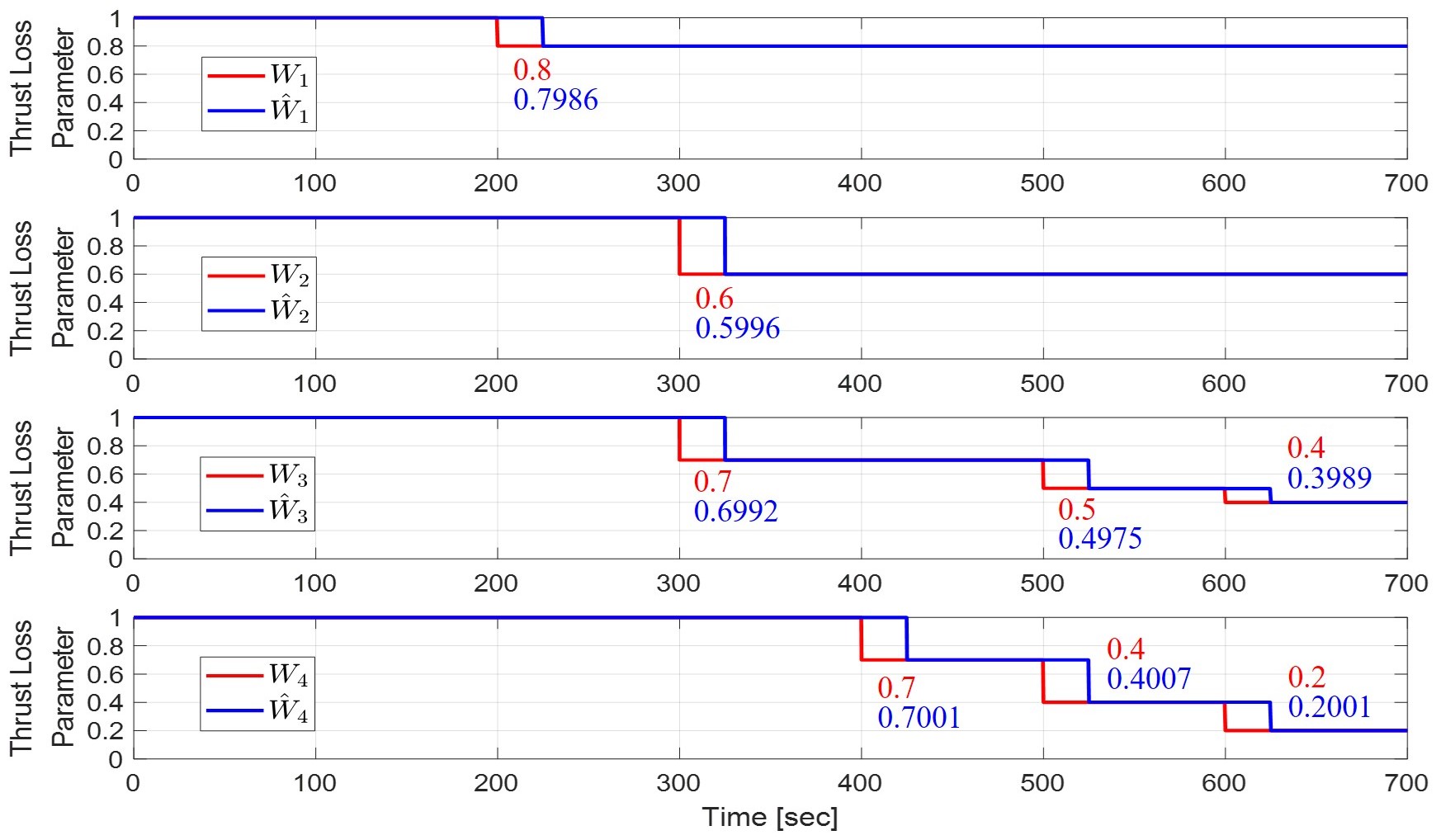}}
\caption{Detection, identification, and thrust loss estimation for single and double thruster fault combinations.}
\label{fig8}
\end{figure}

The corresponding case study results are shown in Fig. \ref{fig8}, demonstrating that the proposed scheme effectively detects and identifies both single and double thruster faults while maintaining accurate estimation performance. In fact, we carried out a large number of similar case studies and observed consistent results throughout.

\subsubsection{Fault Detection Failure}
\begin{figure}[!t]
\centerline{\includegraphics[width=\columnwidth]{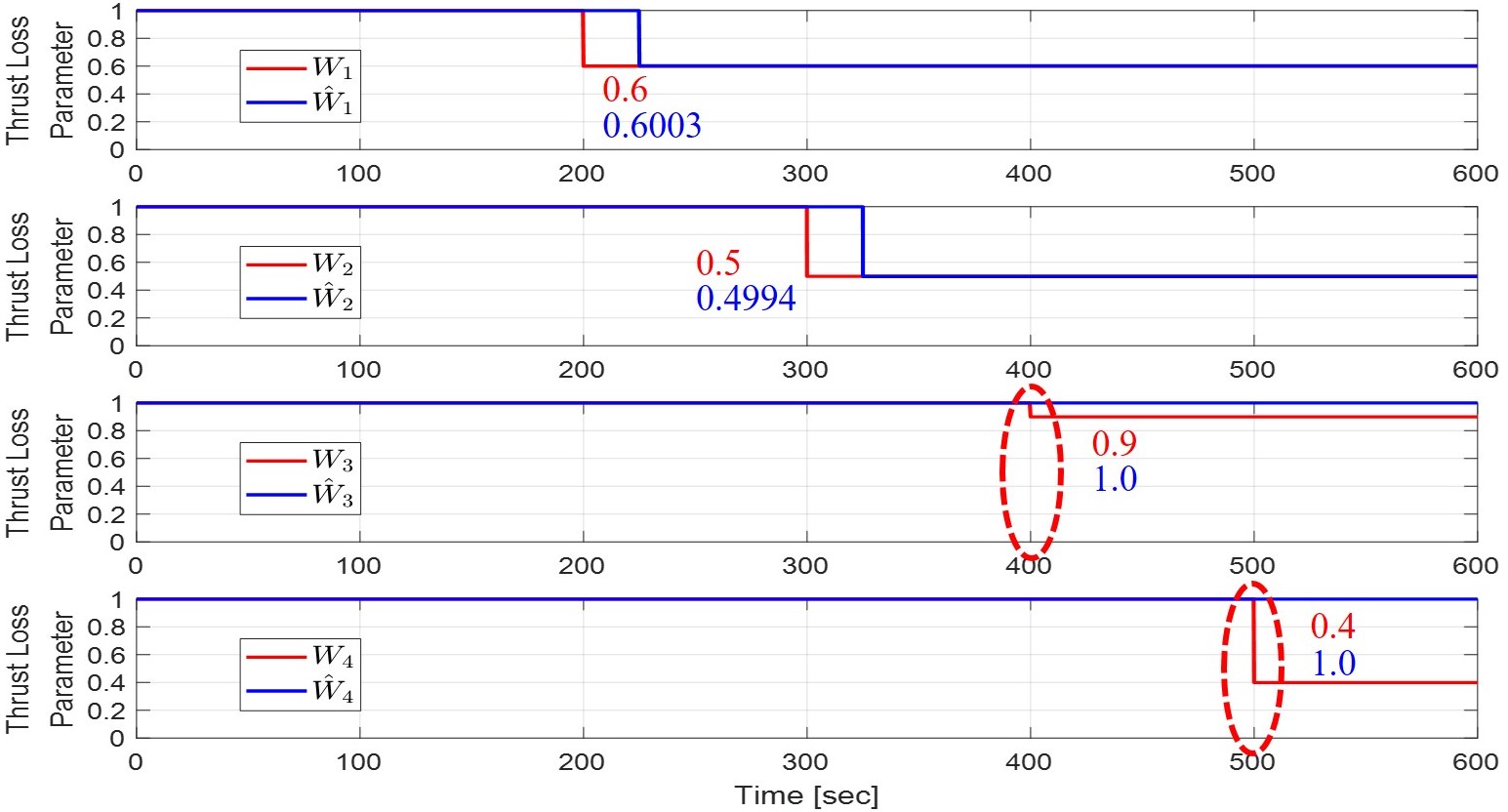}}
\caption{Failure of fault detection in the case of a small thrust loss in the thruster, which corresponds to the third column vector of $\bm{M}_{\tau}$ in (\ref{eq29}), having small magnitude components.}
\label{fig9}
\end{figure}

\begin{figure}[!t]
\centerline{\includegraphics[width=\columnwidth]{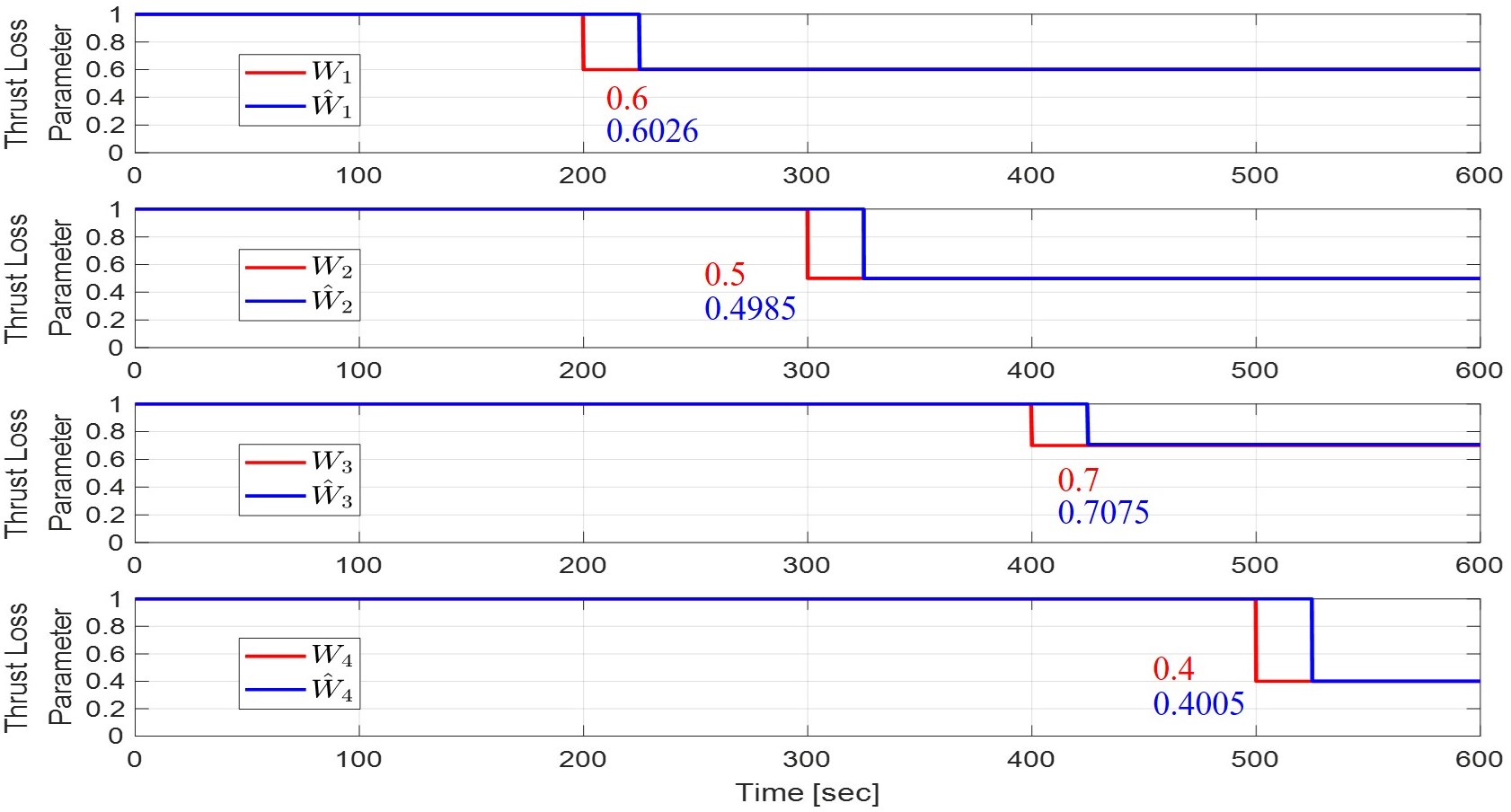}}
\caption{Successful fault detection with increasing thrust loss in the third thruster, while maintaining the same conditions as in Fig. \ref{fig9}.}
\label{fig10}
\end{figure}

If we set $\bm{\eta}_d=[10,2,90^o]^T$ instead of $\bm{\eta}_d=[10,2,70^o]^T$, then $\bm{M}_{\tau}$ takes the following form:
\begin{equation}
\bm{M}_{\tau}=\begin{bmatrix}3.5178&5.0833&-0.7511&9.3522\\-3.5178&5.0833&-0.7511&-9.3522\\-0.9393&1.3572&0.2006&2.4971\end{bmatrix}, \label{eq29}
\end{equation}
where the third column has relatively small component magnitude compared to the others. In this case, if we set $\delta W_3=0.1$, indicating a relatively small thrust loss for the third thruster, Fig. \ref{fig9} shows that the proposed FDI algorithm fails to detect this fault, which further causes the detection failure of $W_4$ at $t=500s$. However, when the thrust loss is increased to $\delta W_3=0.3$, both $\delta W_3$ and $\delta W_4$ becomes detectable, see Fig. \ref{fig10}. As mentioned in Remark 5, in the specific case of $\bm{\tau}_c$ which causes one of $\tau_i$, $i=1,\cdots,4$ to be equal or near zero, the thrust loss in this $i$th thruster becomes difficult to be detected. This could be recognized as an inherent limitation of the FDI algorithm proposed in this paper.

\section{Conclusion}
This paper has presented a novel FTC scheme for a class of underwater vehicles with thruster redundancy. Rather than relying on estimation errors, the residual is constructed by directly applying MCE, eliminating the need for an additional estimation system. Through detailed and precise analyses of MCE variation trends in the presence of thruster faults, valuable information regarding the actual control input after a thruster faults can be acquired, allowing us to formulate both the fault identification method and the thrust loss estimations. The proposed FDI algorithm can handle up to two simultaneous thruster faults. Various numerical studies have been carried out, demonstrating that the proposed method provides satisfactory FDI performance and accurate thrust loss estimation.

Since the thruster configuration matrix is $3\times4$, in principle, simultaneous faults in three thrusters cannot be distinguished from a fault in the remaining single thruster. However, by restricting each thrust loss to the domain $[0,1]$, it may be possible to solve this triple-fault problem. Therefore, this will be an interesting topic for our future research.


\end{document}